\documentclass[twocolumn]{aastex631}
\usepackage{CJK}

\let\tablenum\relax
\usepackage{hyperref}

\usepackage[separate-uncertainty,multi-part-units = single,group-digits=false]{siunitx}
\usepackage{bm}

\usepackage{natbib}

\usepackage{chemformula}
\usepackage{amsmath}
\usepackage{float}

\setcitestyle{notesep={ }}
\shorttitle{JWST Imaging of AF Lep b}
\shortauthors{Franson et al.}

\graphicspath{{./}{figures/}}

\begin{document}
\begin{CJK*}{UTF8}{gbsn}

\title{JWST/NIRCam $\mathbf{4{-}5\, \mathbf{\mu m}}$ Imaging of the Giant Planet AF Lep b}

\author[0000-0003-4557-414X]{Kyle Franson}
\altaffiliation{NSF Graduate Research Fellow}
\affiliation{Department of Astronomy, The University of Texas at Austin, Austin, TX 78712, USA}

\author[0000-0001-6396-8439]{William O. Balmer}
\altaffiliation{Johns Hopkins University George Owen Fellow}
\affiliation{Department of Physics \& Astronomy, Johns Hopkins University, 3400 N. Charles Street, Baltimore, MD 21218, USA}
\affiliation{Space Telescope Science Institute, 3700 San Martin Dr, Baltimore, MD 21218, USA}

\author[0000-0003-2649-2288]{Brendan P. Bowler}
\affiliation{Department of Astronomy, The University of Texas at Austin, Austin, TX 78712, USA}

\author[0000-0003-3818-408X]{Laurent Pueyo}
\affiliation{Space Telescope Science Institute, 3700 San Martin Dr, Baltimore, MD 21218, USA}

\author[0000-0003-2969-6040]{Yifan Zhou}
\affiliation{Department of Astronomy, University of Virginia, 530 McCormick Road, Charlottesville, VA 22904, USA}

\author[0000-0003-4203-9715]{Emily Rickman}
\affiliation{European Space Agency (ESA), ESA Office, Space Telescope Science Institute, 3700 San Martin Dr, Baltimore, MD 21218, USA}

\author[0000-0002-3726-4881]{Zhoujian Zhang (张周健)}
\altaffiliation{NASA Sagan Fellow}
\affiliation{Department of Astronomy \& Astrophysics, University of California, Santa Cruz, CA 95064, USA }

\author[0000-0003-1622-1302]{Sagnick Mukherjee}
\affiliation{Department of Astronomy \& Astrophysics, University of California, Santa Cruz, CA 95064, USA}

\author[0000-0001-5653-5635]{Tim D. Pearce}
\affiliation{Department of Physics, University of Warwick, Gibbet Hill Road, Coventry, CV4 7AL, UK}

\author[0000-0001-8170-7072]{Daniella C. Bardalez Gagliuffi}
\affiliation{Department of Physics \& Astronomy, Amherst College, 25 East Drive, Amherst, MA 01003, USA}
\affiliation{American Museum of Natural History, 200 Central Park West, New York, NY 10024, USA}

\author[0000-0003-2646-3727]{Lauren I. Biddle}
\affiliation{Department of Astronomy, The University of Texas at Austin, Austin, TX 78712, USA}

\author{Timothy D. Brandt}
\affiliation{Space Telescope Science Institute, 3700 San Martin Dr, Baltimore, MD 21218, USA}

\author[0000-0001-5831-9530]{Rachel Bowens-Rubin}
\affiliation{Department of Astronomy \& Astrophysics, University of California, Santa Cruz, CA 95064, USA}

\author[0000-0003-0800-0593]{Justin R. Crepp}
\affiliation{Department of Physics, University of Notre Dame, 225 Nieuwland Science Hall, Notre Dame, IN 46556, USA}

\author[0009-0007-1284-7240]{James W. Davidson, Jr.}
\affiliation{Department of Astronomy, University of Virginia, 530 McCormick Road, Charlottesville, VA 22904, USA}

\author[0000-0001-6251-0573]{Jacqueline Faherty}
\affiliation{American Museum of Natural History, 200 Central Park West, New York, NY 10024, USA}

\author[0000-0002-4438-1971]{Christian Ginski}
\affiliation{School of Natural Sciences, Center for Astronomy, University of Galway, Galway, H91 CF50, Ireland}

\author[0000-0003-2159-1463]{Elliott P. Horch}
\altaffiliation{Adjunct Astronomer, Lowell Observatory}
\affiliation{Department of Physics, Southern Connecticut State University, 501 Crescent Street, New Haven, CT 06515, USA}

\author[0000-0003-4022-6234]{Marvin Morgan}
\affiliation{Department of Astronomy, The University of Texas at Austin, Austin, TX 78712, USA}

\author[0000-0002-4404-0456]{Caroline V. Morley}
\affiliation{Department of Astronomy, The University of Texas at Austin, Austin, TX 78712, USA}

\author[0000-0002-3191-8151]{Marshall D. Perrin}
\affiliation{Space Telescope Science Institute, 3700 San Martin Dr, Baltimore, MD 21218, USA}

\author[0000-0002-1838-4757]{Aniket Sanghi}
\altaffiliation{NSF Graduate Research Fellow}
\affiliation{Department of Astronomy, California Institute of Technology, 1200 E. California Boulevard, Pasadena, CA 91125, USA}

\author[0000-0002-5082-6332]{Ma\"\i ssa Salama}
\affiliation{Department of Astronomy \& Astrophysics, University of California, Santa Cruz, CA 95064, USA}

\author[0000-0002-9807-5435]{Christopher A. Theissen}
\affiliation{Center for Astrophysics and Space Sciences, University of California, San Diego, 9500 Gilman Drive, La Jolla, CA 92093, USA}

\author[0000-0001-6532-6755]{Quang H. Tran}
\affiliation{Department of Astronomy, The University of Texas at Austin, Austin, TX 78712, USA}

\author[0000-0002-1406-8829]{Trevor N. Wolf}
\affiliation{Department of Aerospace Engineering and Engineering Mechanics, The University of Texas at Austin, Austin, TX 78712, USA}
 
\begin{abstract}
    With a dynamical mass of $3 \, M_\mathrm{Jup}$, the recently discovered giant planet AF Lep b is the lowest-mass imaged planet with a direct mass measurement. Its youth and spectral type near the L/T transition make it a promising target to study the impact of clouds and atmospheric chemistry at low surface gravities. In this work, we present JWST/NIRCam imaging of AF Lep b. Across two epochs, we detect AF Lep b in F444W ($4.4 \, \mathrm{\mu m}$) with S/N ratios of $9.6$ and $8.7$, respectively. At the planet's separation of $320 \, \mathrm{mas}$ during the observations, the coronagraphic throughput is ${\approx}7\%$, demonstrating that NIRCam's excellent sensitivity persists down to small separations. The F444W photometry of AF Lep b affirms the presence of disequilibrium carbon chemistry and enhanced atmospheric metallicity. These observations also place deep limits on wider-separation planets in the system, ruling out $1.1 \, M_\mathrm{Jup}$ planets beyond $15.6 \, \mathrm{au}$ (0\farcs58), $1.1 \, M_\mathrm{Sat}$ planets beyond $27 \, \mathrm{au}$ ($1''$), and $2.8 \, M_\mathrm{Nep}$ planets beyond $67 \, \mathrm{au}$ (2\farcs5). We also present new Keck/NIRC2 $L'$ imaging of AF Lep b; combining this with the two epochs of F444W photometry and previous Keck $L'$ photometry provides limits on the long-term 3--$5 \, \mathrm{\mu m}$ variability of AF Lep b on months-to-years timescales. AF Lep b is the closest-separation planet imaged with JWST to date, demonstrating that planets can be recovered well inside the nominal (50\% throughput) NIRCam coronagraph inner working angle.

\end{abstract}

\section{Introduction}
\end{CJK*}

The atmospheres of gas giants and brown dwarfs undergo dramatic changes as they cool and evolve. Thick clouds suspended in the atmospheres of warmer L-dwarfs break up and sink below the photosphere at the L/T transition, yielding the clear atmospheres of cooler T dwarfs \citep{tsujiEvolutionDustyPhotospheres_1996,burrowsTheoryBrownDwarfs_2001,knappNearinfraredPhotometrySpectroscopy_2004}. Coinciding with this transition, near-infrared (NIR) colors shift blueward and the dominant carbon-bearing molecule shifts from carbon monoxide (CO) to methane ($\mathrm{CH_4}$), causing the characteristic $\mathrm{CH_4}$ absorption features of T dwarfs \citep{oppenheimerInfraredSpectrumCool_1995,kirkpatrickNewSpectralTypes_2005}. For field objects, the L/T transition occurs at $T_\mathrm{eff} \sim 1200{-}1400 \, \mathrm{K}$ \citep{filippazzoFundamentalParametersSpectral_2015,sanghiHawaiiInfraredParallax_2023}. However, young, low-gravity giant planets and brown dwarfs follow a different path, retaining clouds, red NIR colors, and CO absorption to much lower effective temperatures than their field-age brown dwarf counterparts \citep[e.g.,][]{chauvinGiantPlanetCandidate_2004,maroisDirectImagingMultiple_2008,bowlerNearinfraredSpectroscopyExtrasolar_2010,bowlerPlanetsLowmassStars_2013,fahertyPopulationPropertiesBrown_2016,liuHawaiiInfraredParallax_2016}. The retention of photospheric condensates may be caused by the dependence of cloud base pressure and particle size with surface gravity \citep{marleyMassesRadiiCloud_2012}, while the preservation of $\mathrm{CO}$ and lack of $\mathrm{CH_4}$ at NIR wavelengths points to disequilibrium carbon chemistry caused by strong vertical mixing in low-surface-gravity atmospheres \citep[e.g.,][]{hinzThermalInfraredMmtao_2010,barmanCloudsChemistryAtmosphere_2011,konopackyDetectionCarbonMonoxide_2013,skemerFirstLightLbt_2012,skemerDirectlyImagedLt_2014}. However, the dearth of imaged planets spanning the L/T transition limits understanding this change in detail.

AF Lep b is a recent addition to the census of young imaged planets. It was independently discovered by three groups searching for planets around stars with astrometric accelerations between Hipparcos and Gaia \citep{fransonAstrometricAccelerationsDynamical_2023a,derosaDirectImagingDiscovery_2023,mesaAfLepLowest_2023}. With a mass of $2.8^{+0.6}_{-0.5} \, M_\mathrm{Jup}$ and semi-major axis of $8.2^{+1.3}_{-1.7} \, \mathrm{au}$ \citep{zhangElementalAbundancesPlanets_2023}, AF Lep b is the lowest-mass directly imaged planet with a dynamical mass measurement. The system is a member of the $\beta$ Pic moving group with an age of $24 \pm 3 \, \mathrm{Myr}$ \citep{bellSelfconsistentAbsoluteIsochronal_2015}. It resides at the L/T transition, with a luminosity between that of the early-L dwarf $\beta$~Pic~b \citep{lagrangeGiantPlanetImaged_2010,bonnefoyPhysicalOrbitalProperties_2014} and the mid-to-late T dwarf 51~Eri~b \citep{macintoshDiscoverySpectroscopyYoung_2015,rajanCharacterizing51Eri_2017}. There are signs of methane absorption in the $K$-band spectral shape of AF Lep b \citep{derosaDirectImagingDiscovery_2023} and evidence for an enhancement in atmospheric metallicity \citep{zhangElementalAbundancesPlanets_2023,palma-bifaniAtmosphericPropertiesAf_2024}, similar to Jupiter and Saturn \citep{atreyaCompositionOriginAtmosphere_2003,flasarTemperaturesWindsComposition_2005}. Based on AF Lep b's bolometric luminosity and dynamical mass, \citet{fransonAstrometricAccelerationsDynamical_2023a}, \citet{zhangElementalAbundancesPlanets_2023}, and \citet{zhangInitialEntropyPotential_2024} found hints of differential ages between the planet and its host star, which may represent the first direct evidence of delayed formation by a few Myr as would be expected for a planet forming through core accretion. The system also hosts an unresolved debris disk with an estimated separation of $46 \pm 9 \, \mathrm{au}$ \citep{pawellek75CentOccurrence_2021, pearcePlanetPopulationsInferred_2022}. Using dynamical arguments, \citet{pearcePlanetPopulationsInferred_2022} found that a $1.1 \pm 0.2 \, M_{\mathrm{Jup}}$ planet at $35 \pm 6 \, \mathrm{au}$ is consistent with truncating the inner edge of this debris disk---perhaps a sign of an outer planet beyond AF Lep b in the system---although it could also be explained by multiple small planets or by AF Lep b forming at wider separations and migrating inward.

Here we present the results of a Director's Discretionary Time program (Program ID 4558; Co-PIs Franson, Balmer) to image AF Lep b with the Near-Infrared Camera \citep[NIRCam;][]{riekeOverviewJamesWebb_2005,riekePerformanceNircamJwst_2023} on JWST \citep{gardnerJamesWebbSpace_2006,gardnerJamesWebbSpace_2023}. At its current separation of ${\sim}320 \, \mathrm{mas}$ ($2.3 \lambda / D$ in F444W) and contrast of ${\sim}10 \, \mathrm{mag}$ in thermal wavelengths \citep{fransonAstrometricAccelerationsDynamical_2023a}, AF Lep b is a challenging object to recover with JWST. However, the recent successful detection of HR 8799 e with the Mid-infrared Instrument \citep[MIRI;][]{boccalettiImagingDetectionInner_2023} and outstanding coronagraphic performance of NIRCam \citep[e.g.,][]{girardJwstNircamCoronagraphy_2022,kammererPerformanceNearinfraredHighcontrast_2022,kammererJwsttstHighContrast_2024,carterJwstEarlyRelease_2023,lawsonJwstNircamCoronagraphy_2023} open the possibility of high-contrast imaging below the nominal (50\% throughput) coronagraph inner working angle. Extending the Spectral Energy Distribution (SED) of AF Lep b to $4{-}5 \, \mathrm{\mu m}$ provides an opportunity to search for signs of disequilibrium chemistry and metallicity enhancement by sampling the $4.3 \, \mathrm{\mu m}$ $\mathrm{CO_2}$ and $4.6 \, \mathrm{\mu m}$ CO absorption bands. Additionally, NIRCam imaging enables a deep search for other planets in the system at wider separations including a potential sculptor of the debris disk. Finally, we also report new Keck/NIRC2 $L'$ imaging as part of a search for variability from 3--$5 \, \mathrm{\mu m}$ over a baseline of 2.1 years.

\begin{figure*}
    \centering
    \includegraphics[width=1\textwidth]{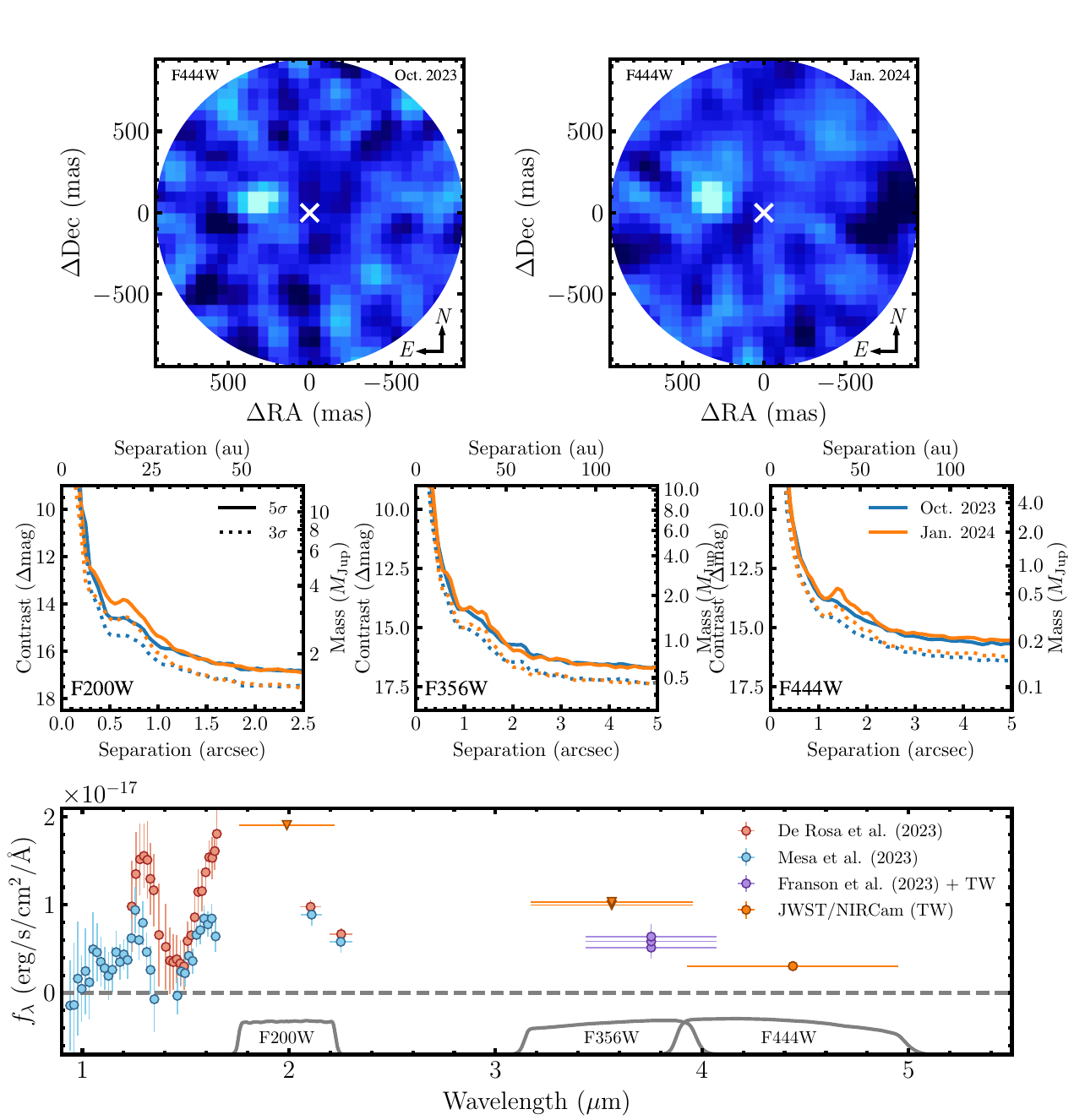}
    \caption{\emph{Top:} JWST/NIRCam imaging of AF Lep b in F444W. Each image is oriented so north is up and east is to the left. To average over pixel-to-pixel noise, the images are convolved with a Gaussian filter with a standard deviation of 1 pixel. AF Lep b is recovered with a S/N of $9.6\sigma$ in October 2023 and $8.7\sigma$ in January 2024 at $4.44\, \mathrm{\mu m}$. \emph{Middle:} $5\sigma$ (solid) and $3\sigma$ (dotted) contrast curves in F200W (left), F356W (middle), and F444W (right). Blue lines denote contrast curves for the October 2023 observations, while orange lines correspond to the January 2024 observations. Each right axis shows corresponding model-inferred masses from ATMO-2020 and \citet{linderEvolutionaryModelsCold_2019}. F444W provides the deepest mass limits, ruling out Saturn-mass planets beyond $1^{\prime\prime}$ ($27 \, \mathrm{au}$) and $0.14 \, M_\mathrm{Jup} = 2.5 \, M_\mathrm{Nep}$ planets beyond $2\farcs5$ ($67 \, \mathrm{au}$). \emph{Bottom:} Spectral energy distribution of AF Lep b from $1{-}5 \, \mathrm{\mu m}$. SPHERE IRDIS photometry and IFS spectroscopy from \citet{derosaDirectImagingDiscovery_2023} and \citet{mesaAfLepLowest_2023} are shown in red and blue, respectively. Keck/NIRC2 $L'$ photometry from \citet{fransonAstrometricAccelerationsDynamical_2023a} and this work are shown in purple. Our new JWST/NIRCam F444W photometry is shown in orange with $5\sigma$ upper limits denoted by upside-down triangles. Note that the January 2024 F200W upper limit of $2.8 \times 10^{-17} \, \mathrm{erg/s/cm^2/\text{\AA}}$ is outside the bounds of this figure.}

    \label{fig:reducedimgs}
\end{figure*}

\vskip 0.1 in

\section{Observations and Data Reduction}
\subsection{JWST/NIRCam Imaging \label{sec:jwst_methods}}
We obtained F200W ($1.755{-}2.227 \, \mathrm{\mu m}$), F356W ($3.135{-}3.981 \, \mathrm{\mu m}$), and F444W ($3.881{-}4.982 \, \mathrm{\mu m}$) JWST/NIRCam imaging of AF Lep on UT 2023 October 12. This dataset was affected by a severe mirror ``tilt event" shortly before the observations. Tilt events are occassional abrupt changes in the position of one or more mirror segments thought to arise from structural microdynamics within the telescope \citep{rigbySciencePerformanceJwst_2023}. The wavefront stability is measured with NIRCam approximately every two days \citep{actonWavefrontSensingControls_2012,mcelwainJamesWebbSpace_2023}. At UT 2023 October 12 06:53:15.4, 14 minutes before the imaging sequence, NIRCam measured a total wavefront error (WFE) of $96.2\, \mathrm{nm}$ rms, compared to a baseline of $69.2 \, \mathrm{nm}$ rms on UT 2023 October 5. At the time, this was the highest WFE since October 2022 and the second highest since commissioning. Because of the potential impact of this tilt event on the dataset, a second set of observations were obtained on UT January 12 2024. These observations were acquired under nominal wavefront conditions: a WFE of 67.2 nm rms was measured on UT 2024 January 12 19:55:41.5 and 66.2 nm rms on UT 2024 January 15 10:23:19.7. Despite the tilt event, we elect to analyze both datasets in this work because the October sequence retains two advantages over the January data: AF Lep b is at a slightly wider separation and the star is better centered behind the coronagraph, as detailed below.

Each sequence consists of two roll angles centered on AF Lep immediately followed by deep imaging of the reference star HD 33093. This approach facilitates both Angular Differential Imaging \citep[ADI;][]{liuSubstructureCircumstellarDisk_2004,maroisAngularDifferentialImaging_2006} and Reference Star Differential Imaging \citep[RDI;][]{lafreniereHstNicmosDetection_2009}. HD 33093 was selected as the reference star for this program based on its similar magnitude, spectral type, and angular proximity to AF Lep.\footnote{HD 33093 has a spectral type of G0IV \citep{grayContributionsNearbyStars_2006} and $W1$-band magnitude of $4.43 \pm 0.09 \, \mathrm{mag}$ \citep{cutriVizierOnlineData_2012}. This closely matches the spectral type \citep[F8V;][]{grayContributionsNearbyStars_2006} and magnitude \citep[$W1 {=} 4.92 \pm 0.07 \, \mathrm{mag}$;][]{cutriVizierOnlineData_2012} of AF Lep. The angular separation between the stars is 4\fdg8. HD 33093 also has a Renormalised Unit Weght Error (RUWE; \citealt{lindegrenRenormalisingAstrometricChisquare_2018}) in Gaia DR3 of 1.047, fraction of double transits parameter (IFDsp; see e.g., \citealt{tokovininExploringThousandsNearby_2023}) of 0, and lacks a significant astrometric acceleration in the Hipparcos-Gaia Catalog of Accelerations (HGCA; \citealt{brandtHipparcosgaiaCatalogAccelerations_2021}), all of which point to the star being single.}\footnote{To ensure that the reference star is not a close visual binary, we obtained diffraction-limited imaging of HD 33093 on UT 2023 September 1 using the Differential Speckle Survey Instrument \citep[DSSI,][]{horchObservationsBinaryStars_2009}, a visible-light speckle camera currently on the Astrophysical Research Consortium 3.5-m telescope at Apache Point Observatory \citep[See][]{davidsonObservationsDifferentialSpeckle_2024}. HD 33093 was also observed with VLT/SPHERE \citep{beuzitSphereExoplanetImager_2019} on UT 2023 October 16. The star appeared single in both datasets.} Each epoch is executed as a non-interrutible sequence to minimize wavefront drift. The roll angles are separated by $10.0^\circ$ in each sequence. All imaging is taken with the \texttt{335R} occulting mask \citep{kristJwstNircamCoronagraph_2010}, which has a smooth transmission function with an Inner Working Angle (IWA) of 0\farcs63. To ensure a sufficient diversity of reference PSFs and to mitigate the impact of target acquisition errors, the \texttt{9-POINT-CIRCLE} dither pattern is used for the reference star observations. F200W imaging is taken simultaneously with F356W and F444W imaging using the NIRCam dichroic. We use the \texttt{SUB320A335R} subarray to minimize readout times. This gives a field of view of $20^{\prime\prime}\times20^{\prime\prime}$ in F356W/F444W and $10^{\prime\prime}\times10^{\prime\prime}$ in F200W. In each sequence, we obtained 35 integrations of five groups in the \texttt{SHALLOW4} readout pattern for each filter combination (F200W/F356W and F200W/F444W) and roll angle of AF Lep. For the first epoch in October 2023, the reference star observations consist of 14 integrations in F200W/F444W and 13 integrations in F200W/F356W of five \texttt{SHALLOW4} groups per dither position. For the second sequence in January 2024, 15 integrations of the reference star are obtained in each dither position and filter combination. For both sequences, the total exposure time on AF Lep is 62.4 min in F200W and 31.2 min in F356W and F444W. The total exposure time on the reference star in the October 2023 observations is 108.3 min in F200W, 52.2 min in F356W, and 56.2 min in F444W. For the January 2024 observations, the reference star exposure time is 120.4 min in F200W and 60.2 min in both F356W and F444W.

We process uncalibrated Stage 0 files into Stage 1 and Stage 2 data products using \texttt{spaceKLIP}\footnote{We use version 2 of the development branch of \texttt{spaceKLIP} (commit \#f64258d
).} \citep{kammererPerformanceNearinfraredHighcontrast_2022}, which makes use of the \texttt{jwst} pipeline\footnote{We use pipeline version $\mathtt{CAL\_VAR} = 1.12.5$ and calibration reference data jwst\_1183.pmap.} \citep{bushouseJwstCalibrationPipeline_2023} for basic data processing steps with modifications for coronagraphic imaging reduction as detailed in \citet{carterJwstEarlyRelease_2023}. Following \citet{carterJwstEarlyRelease_2023}, we do not carry out dark current subtraction, as the NIRCam dark current calibration data contains a large number of hot pixels, cosmic rays, and persistence features that result in reduced sensitivity when included. We use a jump detection threshold of four when reducing the up-the-ramp uncalibrated data. In this step \texttt{spaceKLIP} applies a $1/f$ noise correction: a master slope image is subtracted from each ramp image to produce a residual that is modeled using a Savitzky-Golay filter. This model (which approximates the $1/f$ noise) is then removed from each ramp frame (see e.g., \citealt{rebollidoJwsttstHighContrast_2024}). Following processing into Stage 2 calibrated products, we perform standard pixel cleaning procedures for NIRCam high-contrast imaging. Pixels flagged as poor data quality by the \texttt{jwst} pipeline together with $3\sigma$ outliers are replaced with the median of neighboring pixels. We also identify anomalous pixels by flagging pixels with significant temporal flux variations between integrations. Pixels brighter than $1 \, \mathrm{MJy\,sr^{-1}}$ or fainter than $20\%$ of the brightest value for a given pixel position are replaced with the median value of the pixel in the exposure sequence. The position of the star behind the coronagraph is determined by fitting a model coronagraphic PSF from \texttt{webbpsf\_ext}.\footnote{\url{https://github.com/JarronL/webbpsf_ext}} We find that AF Lep is better centered in the October 2023 imaging (${\approx} 1.5$ mas offset) than the January 2024 imaging (${\approx}20$ mas offset). The January 2024 centering offset is large but not an abnormal deviation compared to the target acquisition performance of NIRCam with the \texttt{335R} mask ($\sigma \approx 15 \, \mathrm{mas}$\footnote{\href{https://jwst-docs.stsci.edu/jwst-calibration-status/nircam-calibration-status/nircam-coronagraphy-calibration-status}{https://jwst-docs.stsci.edu/jwst-calibration-status/nircam-calibration-status/nircam-coronagraphy-calibration-status}}).

We carry out PSF subtraction with \texttt{spaceKLIP}, which uses the Karhunen--Lo\`{e}ve Image Projection \citep[KLIP;][]{soummerDetectionCharacterizationExoplanets_2012} algorithm implemented in \texttt{pyKLIP} \citep{wangPyklipPsfSubtraction_2015} to model and subtract the host-star PSF. The following combination of parameters effectively suppresses residual speckle noise for each filter: 100 KL modes, four annuli, and three subsections, utilizing both ADI and RDI for reference frames. Figure \ref{fig:reducedimgs} shows the final PSF-subtracted images in F444W.\footnote{See Appendix \ref{sec:all_imgs} for F356W and F200W imaging.} AF Lep b is detected in F444W at S/N ratios of $9.6$ and $8.7$ for the October 2023 and January 2024 epochs, respectively. Here, S/N is measured by comparing the flux in a circular aperture at the position of AF Lep b against the noise level estimated by ten non-overlapping apertures at the same separation. To correct for small-sample statistics at small separations, we adopt the Student's t-distribution with equivalent false-alarm probabilities to our Gaussian significance levels following \citet{mawetFundamentalLimitationsHigh_2014}. The comparable detection significance between the two datasets despite the higher WFE in the October 2023 dataset may be due to the larger centering offset in the January 2024 sequence. AF Lep b is not recovered at a significant level (above $5\sigma$) in other filters. No other significant sources are detected in the full-frame images.\footnote{An artifact appears in the October 2023 F444W reduced images at a separation of 5\farcs7 and position angle of $27^\circ$. This appears to be caused by a cosmic ray hitting the detector between rolls, causing a ``snowball'' artifact that is missed by jump detection \citep[see e.g.,][]{bagleyCeersEpochNircam_2023}. The artifact is removed when the first five integrations of the second roll angle are excluded.}

To measure the astrometry and photometry of AF Lep b, we use the KLIP forward modeling \citep[KLIP-FM;][]{pueyoDetectionCharacterizationExoplanets_2016} framework implemented in \texttt{spaceKLIP} and \texttt{pyKLIP}. This approach yields a flexible forward model that incorporates the distortions that KLIP introduces to a planet's PSF. This can then be fit to a given source in a post-processed image within a Bayesian framework \citep[see e.g.,][]{wangOrbitTransitProspects_2016}. We employ the \texttt{emcee} affine-invariant Markov-chain Monte Carlo (MCMC) ensemble sampler \citep{foreman-mackeyEmceeMcmcHammer_2013} to simultaneously fit the planet separation, position angle, and flux ratio. The PSF model is produced by \texttt{webbpsf} \citep{perrinSimulatingPointSpread_2012,perrinUpdatedPointSpread_2014}. A total of 50 walkers over $1.5 \times 10^{4}$ total steps (300 per walker) are used to sample the three model parameter posteriors. We discard the first 100 steps of each chain as burn-in. As speckle noise is correlated at the separation of AF Lep b, we model the noise distribution via a Gaussian process with a Mat{\'e}rn ($\nu = 3/2$) kernel following \citet{wangOrbitTransitProspects_2016}. This increases the photometric uncertainties from ${\approx}{\pm}0.10 \, \mathrm{mag}$ to ${\approx}{\pm}0.15 \, \mathrm{mag}$ compared to assuming uncorrelated noise (the ``diagonal kernel'' in \texttt{spaceKLIP}).

Astrometric and photometric uncertainties are determined from the standard deviation of the MCMC chains. Following \citet{carterJwstEarlyRelease_2023}, the adopted astrometric measurements incorporate a 6.3 mas (0.1 pixel) absolute centering uncertainty added in quadrature to account for the precision of the measurement of the host-star position behind the mask. Additionally, a 1.5\% uncertainty is added to the contrast uncertainties in quadrature to reflect the absolute flux calibration accuracy of NIRCam.\footnote{\href{https://jwst-docs.stsci.edu/jwst-calibration-status/nircam-calibration-status/nircam-coronagraphy-calibration-status}{https://jwst-docs.stsci.edu/jwst-calibration-status/nircam-calibration-status/nircam-coronagraphy-calibration-status}}\footnote{See \citet{gordonJamesWebbSpace_2022} for details on the JWST absolute flux calibration program.} To incorporate the throughput of the coronagraph, \texttt{spaceKLIP} uses \texttt{webbpsf\_ext} to calculate the throughput at a provided position of the planet relative to the host star. During this procedure, the misalignment between the host star and the mask is also incorporated into the throughput measurements. We determine the best estimate of the true position of AF Lep b at each epoch from an updated \texttt{orbitize!} \citep{bluntOrbitizeComprehensiveOrbitfitting_2020} orbit fit incorporating VLTI/GRAVITY astrometry of AF Lep b (Balmer et al., in prep) and \texttt{whereistheplanet} \citep{wangWhereistheplanetPredictingPositions_2021}. This produces $\rho = 324.17 \pm 0.07 \, \mathrm{mas}$ and $\theta = 78\fdg209 \pm 0\fdg017$ for the October 2023 epoch and $\rho = 319.22 \pm 0.07 \, \mathrm{mas}$ and $\theta = 80\fdg426 \pm 0\fdg012$ for the January 2024 epoch. Incorporating a conservative host star position uncertainty of 0.1 pixels ($6.3 \, \mathrm{mas}$) yields throughput values of $8.4 \pm 0.5$\% (October 2023) and $6.0 \pm 0.4$\% (January 2024). These throughput uncertainties are included in the final photometric uncertainties.
\vskip -0.34 in
\begin{deluxetable*}{lccccccc} 
\tablecaption{\label{tab:results}New Astrometry and Photometry of AF Lep b}
\tablehead{\colhead{Date} & \colhead{Epoch} & \colhead{Instrument} & \colhead{Filter} & \colhead{$\rho$} & \colhead{$\theta$} & \colhead{$\Delta_\mathrm{mag}$} & \colhead{$f_\lambda \times 10^{-18}$}\\
\colhead{(UT)} & \colhead{(UT)} & \colhead{} & \colhead{} & \colhead{(mas)} & \colhead{(\si{\degree})} & \colhead{(mag)} & \colhead{($\mathrm{erg/s/cm^2/\text{\AA}}$)}}
\startdata
2023 Oct 12 & 2023.779 & JWST/NIRCam & F444W & $308 \pm 18$ & $73.6 \pm 2.4$ & $9.98 \pm 0.16$ & $3.0 \pm 0.4$\\
2023 Oct 12 & 2023.779 & JWST/NIRCam & F356W & $. . .$ & $. . .$ & ${>}9.6\tablenotemark{a}$ & ${<}10\tablenotemark{a}$\\
2023 Oct 12 & 2023.779 & JWST/NIRCam & F200W & $. . .$ & $. . .$ & ${>}11.2\tablenotemark{a}$ & ${<}19\tablenotemark{a}$\\
2024 Jan 14 & 2024.038 & JWST/NIRCam & F444W & $313 \pm 16$ & $76.2 \pm 2.7$ & $9.98 \pm 0.18$ & $3.0 \pm 0.5$\\
2024 Jan 14 & 2024.038 & JWST/NIRCam & F356W & $. . .$ & $. . .$ & ${>}9.5\tablenotemark{a}$ & ${<}10\tablenotemark{a}$\\
2024 Jan 14 & 2024.038 & JWST/NIRCam & F200W & $. . .$ & $. . .$ & ${>}10.8\tablenotemark{a}$ & ${<}28\tablenotemark{a}$\\
2024 Jan 30 & 2024.080 & Keck/NIRC2 & $L'$ & $300 \pm 7$ & $80.6 \pm 1.2$ & $9.90 \pm 0.25$ & $6.4 \pm 1.4$
\enddata
\tablenotetext{a}{$5\sigma$ limits measured through injection-recovery at the separation of AF Lep b.}\end{deluxetable*}

The resulting astrometry and photometry are shown in Table \ref{tab:results}. For the October 2023 imaging, we measure a separation $\rho = 308 \pm 18\, \mathrm{mas}$, position angle $\theta = 73\fdg6 \pm 2\fdg4$, and F444W contrast $\Delta_{\mathrm{F444W}} = 9.98 \pm 0.15 \, \mathrm{mag}$. For the January 2024 observations, we measure $\rho = 313 \pm 16 \, \mathrm{mas}$, $\theta = 76\fdg2 \pm 2\fdg7$, and $\Delta_\mathrm{F444W} = 9.98 \pm 0.18 \, \mathrm{mag}$. To convert contrasts into apparent magnitudes and flux densities, we synthesize photometry of the host star by scaling a F8V \citep{grayContributionsNearbyStars_2006} BOSZ \citep{bohlinNewStellarAtmosphere_2017} stellar model to photometry of AF Lep from Gaia DR3 \citep{gaiacollaborationGaiaDataRelease_2022}, Tycho-2 \citep{hogTycho2CatalogueMillion_2000}, 2MASS \citep{skrutskieTwoMicronAll_2006}, and WISE \citep{cutriVizierOnlineData_2012}. This produces apparent magnitudes of $m_\mathrm{F444W} = 14.93 \pm 0.15 \, \mathrm{mag}$ and $m_\mathrm{F444W} = 14.93 \pm 0.18 \, \mathrm{mag}$ and flux densities of $f_\mathrm{F444W} = (3.0 \pm 0.4) \times 10^{-18} \, \mathrm{erg/s/cm^2/\text{\AA}}$ and $f_\mathrm{F444W} = (3.0 \pm 0.5) \times 10^{-18} \, \mathrm{erg/s/cm^2/\text{\AA}}$.

We measure $5\sigma$ upper limits in F200W and F356W through injection-recovery. Synthetic sources generated with \texttt{webbpsf} are injected at the same separation as AF Lep b across ${\approx} 13$ non-overlapping position angles. The S/N of each injected source is then measured by the same method as above, here masking both the position of AF Lep b and the synthetic source. The injected flux is then adjusted to produce a S/N of $5\sigma$. The mean injected flux across all position angles is taken as the $5\sigma$ upper limit. These upper limits are shown in Table \ref{tab:results}. We measure F200W upper limits of ${>}11.2 \, \mathrm{mag}$ and ${>} 10.8 \, \mathrm{mag}$ and F356W upper limits of ${>} 9.6 \, \mathrm{mag}$ and ${>} 9.5 \, \mathrm{mag}$ for the October and Janurary imaging, respectively. These correspond to flux densities of ${<}19 \times 10^{-18} \, \mathrm{erg/s/cm^2/\text{\AA}}$ and ${<}28 \times 10^{-18} \, \mathrm{erg/s/cm^2/\text{\AA}}$ in F200W and ${<}10 \times 10^{-18} \, \mathrm{erg/s/cm^2/\text{\AA}}$ and ${<}10 \times 10^{-18}\, \mathrm{erg/s/cm^2/\text{\AA}}$ in F356W for the two epochs.

We compute $3\sigma$ and $5\sigma$ contrast curves with \texttt{spaceKLIP} for each filter and epoch. Across a range of separations, the noise level is measured in annuli within the PSF-subtracted images while masking the position of AF Lep b. The \citet{mawetFundamentalLimitationsHigh_2014} correction for small-sample statistics at small separations is then applied and the noise level is corrected for coronagraphic and algorithmic throughput. Algorithmic throughput is determined by performing injection-recovery with synthetic PSFs generated with \texttt{webbpsf}. Contrast curves for each filter are shown in the middle row of Figure \ref{fig:reducedimgs}.

\vskip .2in
\subsection{Keck/NIRC2 Imaging}
We obtained high-contrast imaging of AF Lep on UT January 30 2024 with the NIRC2 camera at W.M. Keck Observatory using natural guide star adaptive optics \citep{wizinowichAstronomicalScienceAdaptive_2013}. Our observations are taken in pupil-tracking mode to facilitate ADI. No coronagraph is used for our sequence. Each science frame consists of 100 coadds of an integration time of $0.054 \, \mathrm{s}$. To minimize read-out times, we use a subarray of $512\times512$ pixels ($5\farcs1 \times 5\farcs1$). Throughout the sequence, we also periodically take short frames with a smaller subarray ($192\times248$ pixels) to obtain unsaturated images of the host star for flux calibration. These PSF frames each have 100 coadds of $0.017\, \mathrm{s}$. The total on-source integration time is 90.8 min, amounting to $99.9^\circ$ of frame rotation.

Basic data reduction steps follow \citet{fransonDynamicalMassYoung_2022}. After flat-fielding and subtracting darks, we identify and remove cosmic rays using \texttt{L.A. Cosmic} \citep{vandokkumCosmicRayRejection_2001}. To correct for geometric distortions, we apply the \citet{serviceNewDistortionSolution_2016} solution using \texttt{rain},\footnote{\href{https://github.com/jsnguyen/rain}{https://github.com/jsnguyen/rain}} which implements the \texttt{Drizzle} image recombination algorithm \citep{fruchterDrizzleMethodLinear_2002} in Python. Frames are registered by fitting two-dimensional Gaussians to determine the center of each PSF and then all images are aligned to the sub-pixel level using a third-order spline. PSF subtraction is carried out using \texttt{pyKLIP} \citep{wangPyklipPsfSubtraction_2015} with the following parameters: 20 KL modes, three annuli, four subsections, and a movement parameter of one. We recover AF Lep b at a S/N of $6.6\sigma$. 

Astrometry and photometry are measured with \texttt{pyKLIP} using the KLIP-FM framework \citep{pueyoDetectionCharacterizationExoplanets_2016,wangOrbitTransitProspects_2016}. We use \texttt{emcee} \citep{foreman-mackeyEmceeMcmcHammer_2013} to sample the companion parameter space with 100 walkers and $1.6\times10^5$ total steps (1600 steps per walker). The first 25\% of each chain is discarded as burn-in. Following \citet{fransonDynamicalMassYoung_2022}, astrometric uncertainties incorporate the standard deviation of the separation and position angle posterior distributions, the uncertainty on the \citet{serviceNewDistortionSolution_2016} distortion solution of $\sigma_d = 1 \, \mathrm{mas}$, the plate scale uncertainty, and the north alignment uncertainty added in quadrature. Our contrast uncertainty is taken from the standard deviation of the flux ratio posterior. We measure a separation of $300 \pm 7 \, \mathrm{mas}$, position angle of $80\fdg6 \pm 1\fdg2$, and $L'$ contrast of $9.90 \pm 0.25 \, \mathrm{mag}$. Incorporating the $W1$ magnitude of AF Lep A ($4.92 \pm 0.07$; \citealt{cutriVizierOnlineData_2012}), this contrast corresponds to an apparent magnitude of $14.83 \pm 0.26 \, \mathrm{mag}$, an absolute magnitude of $12.68 \pm 0.26 \, \mathrm{mag}$, and a flux density $f_\lambda = (6.4 \pm 1.4) \times 10^{-18} \, \mathrm{erg / s / cm^2 / \text{\AA}}$.

\section{Results}

\begin{figure*}
    \centering
    \includegraphics[width=1\textwidth]{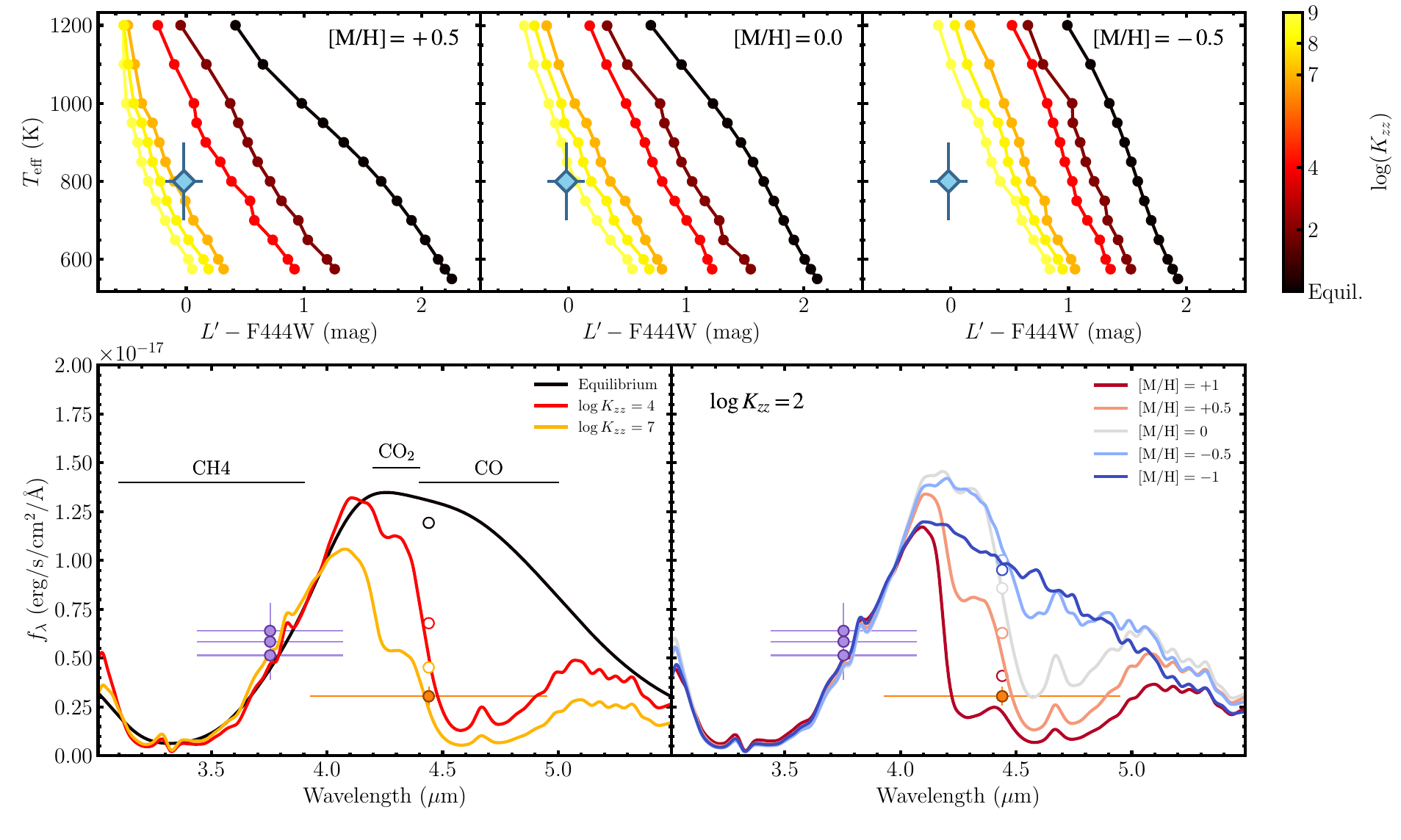}
    \caption{\emph{Top:} $T_\mathrm{eff}$ vs. $L'{-}\mathrm{F444W}$ color of AF Lep b (blue diamond), compared against \texttt{Sonora Elf-Owl} \citep{mukherjeeSonoraSubstellarAtmosphere_2024} and \texttt{Bobcat} \citep{marleySonoraBrownDwarf_2021} model atmospheres, varying metallicity and $\log{K_\mathrm{zz}}$. \texttt{Bobcat} models (black) do not have $\log{K_\mathrm{zz}}$ as a tunable parameter since they assume chemical equilibrium. Here, we include models with $\log g = 3.5 \, \mathrm{dex}$ and solar C/O values. \emph{Bottom:} SED of AF Lep b from $3-5 \, \mathrm{\mu m}$. F444W photometry is shown in orange and $L'$ photometry is shown in purple. The background tracks are \texttt{Elf-Owl} and \texttt{Bobcat} model atmospheres smoothed to $R = 100$ and anchored to the weighted mean of the $L'$ photometry. Open circles show synthesized F444W photometry from the model spectra. The left panel includes models spanning equilibrium chemistry to disequilibrium chemistry with $\log K_{zz} = 7$ and metallicity from $-0.5$ to $+0.5$. The right panel includes \texttt{Elf-Owl} models with $\log{K_{zz}} = 2$ and metallicity spanning a wider range of $-1$ to $+1$. Our F444W photometry is consistent with models that have a high $\log K_{zz}$ (more vigorous mixing) and $\mathrm{[Fe/H]} = 0$ or $+0.5$, or models with lower $\log K_{zz}$ (less vigorous mixing) and a super-solar metallicity ($\mathrm{[Fe/H] = 1}$). Equilibrium and sub-solar models are inconsistent with our photometry.
    \label{fig:combined_analysis}}
\end{figure*}

\subsection{Limits on Outer Planets \label{sec:mass_lims}}
To convert the NIRCam contrast curves to mass limits, we follow \citet{carterDirectImagingSubjupiter_2021} in combining the ATMO-2020 hot-start evolutionary models \citep{phillipsNewSetAtmosphere_2020}, which span 0.5--$75 \, M_\mathrm{Jup}$, and the BEX models \citep{linderEvolutionaryModelsCold_2019}, which span 0.16--$2 \, M_\mathrm{Jup}$. To ensure consistency with the BEX models, which do not incorporate disequilibrium chemistry, the ATMO-2020 grid in chemical equilibrium is adopted. Since ATMO-2020 models are cloudless, we select the cloudless, solar-metallicity BEX grid computed via \texttt{petitCODE} \citep{molliereModelAtmospheresIrradiated_2015}. \citet{linderEvolutionaryModelsCold_2019} computed evolutionary tracks with three prescriptions for the initial luminosity of the planet: a nominal post-formation luminosity based on the output of a population synthesis model of planet formation \citep{mordasiniCharacterizationExoplanetsTheir_2017}, a ``hotter-start" scenario where the initial luminosity is increased by an order of magnitude, and a ``colder-start" scenario where the initial luminosity is decreased by an order of magnitude. We adopt the nominal case, which has the most extensive set of evolutionary tracks. Note that above ages of $1\, \mathrm{Myr}$, the difference between the nominal and hotter-start cases is minimal \citep{linderEvolutionaryModelsCold_2019}. Each model grid is interpolated using the \texttt{species} package \citep{stolkerMiraclesAtmosphericCharacterization_2020} to generate model-inferred masses as a function of contrast within each filter, assuming an age of $24\, \mathrm{Myr}$ for AF Lep based on the $\beta$ Pic moving group age \citep[$24 \pm 3 \, \mathrm{Myr}$;][]{bellSelfconsistentAbsoluteIsochronal_2015}. At contrasts covered by both model grids, we take the average of the two masses.

Figure \ref{fig:reducedimgs} shows our mass limits as a function of separation. In the background-limited regime, we are sensitive to Jupiter-mass planets in F200W ($1.7 \, M_\mathrm{Jup}$ beyond $3^{\prime\prime}$), sub-Jupiter-mass planets in F356W ($0.75 M_\mathrm{Jup}$ beyond $2^{\prime\prime}$), and sub-Saturn-mass planets in F444W ($ 0.2 M_\mathrm{Jup} = 3.6M_\mathrm{Nep}$ beyond $2^{\prime\prime}$). Our F444W imaging is most sensitive to low-mass planets. We rule out planets above masses of $1.9 \, M_\mathrm{Jup}$ at 0\farcs5 ($13 \, \mathrm{au}$), $0.34 \, M_\mathrm{Jup} = 1.1 \, M_\mathrm{Sat}$ at $1^{\prime\prime}$ ($27 \, \mathrm{au}$), $0.20 \, M_\mathrm{Jup} = 0.65 \, M_\mathrm{Sat}$ at $2^{\prime\prime}$ ($67 \, \mathrm{au}$), and $0.15 \, M_\mathrm{Jup} = 0.51 M_\mathrm{Sat} = 2.8 M_\mathrm{Nep}$ from 2\farcs5 to $5^{\prime\prime}$ ($67{-}134 \, \mathrm{au}$). If instead, we assume cold-start formation conditions, the \citet{linderEvolutionaryModelsCold_2019} models produce mass limits of $0.44 \, M_\mathrm{Jup}$, $0.26 \, M_\mathrm{Jup}$, and $0.24 \, M_\mathrm{Jup}$ at $1^{\prime\prime}$, $2^{\prime\prime}$, and $2\farcs5$, respectively. The cold-start grid only spans 0.06--$1\, M_\mathrm{Jup}$; at 0\farcs5, the F444W contrast is outside the bounds of the grid.

AF Lep hosts a debris disk with an estimated separation of $46 \pm 9 \, \mathrm{au}$, inferred from infrared excess \citep{pawellek75CentOccurrence_2021,pearcePlanetPopulationsInferred_2022}, raising the possibility that one or more planets are sculpting its inner edge. Using dynamical arguments, \citet{pearcePlanetPopulationsInferred_2022} determined that the minimum mass for a single giant planet to truncate the debris disk is $1.1 \pm 0.2 \, M_\mathrm{Jup}$ at a semi-major axis of $35 \pm 6 \, \mathrm{au}$. Other possibilities are that the inner edge is set by multiple smaller planets with masses as low as $0.09 \pm 0.03 \, M_\mathrm{Jup}$ or that AF Lep b formed closer to the disk and then migrated inwards. With JWST, we largely rule out the single, non-migrating planet scenario. The F444W imaging is sensitive to planet masses of $1.1 \, M_\mathrm{Jup}$ beyond a projected separation of 0\farcs58 ($15.6 \, \mathrm{au}$) assuming hot-start formation, or $1 \, M_\mathrm{Jup}$ beyond 0\farcs64 ($17.2 \, \mathrm{au}$) in the cold-start scenario. However, a sculpting planet could have been at an unfavorable phase in its orbit during our observations. To estimate the probability of this happening by chance, we employ a Monte Carlo approach by computing $10^4$ circular orbits with isotropic inclinations and semi-major axes drawn from the \citet{pearcePlanetPopulationsInferred_2022} prediction ($35 \pm 6 \, \mathrm{au}$). We find that planets on these orbits are beyond 0\farcs58 95\% of the time. If instead, we assume that the debris disk is coplanar with AF Lep b ($i = 55^{+8}_{-13} {}^\circ$; \citealt{zhangElementalAbundancesPlanets_2023}) and sample planet orbits from that inclination distribution, 90\% of the orbits result in a planet at a projected separation beyond 0\farcs58. The non-detection of an additional planet in the system therefore largely rules out the hypothesis that an additional giant planet is truncating the debris disk around AF Lep.

\begin{figure*}
    \centering
    \includegraphics[width=1\textwidth]{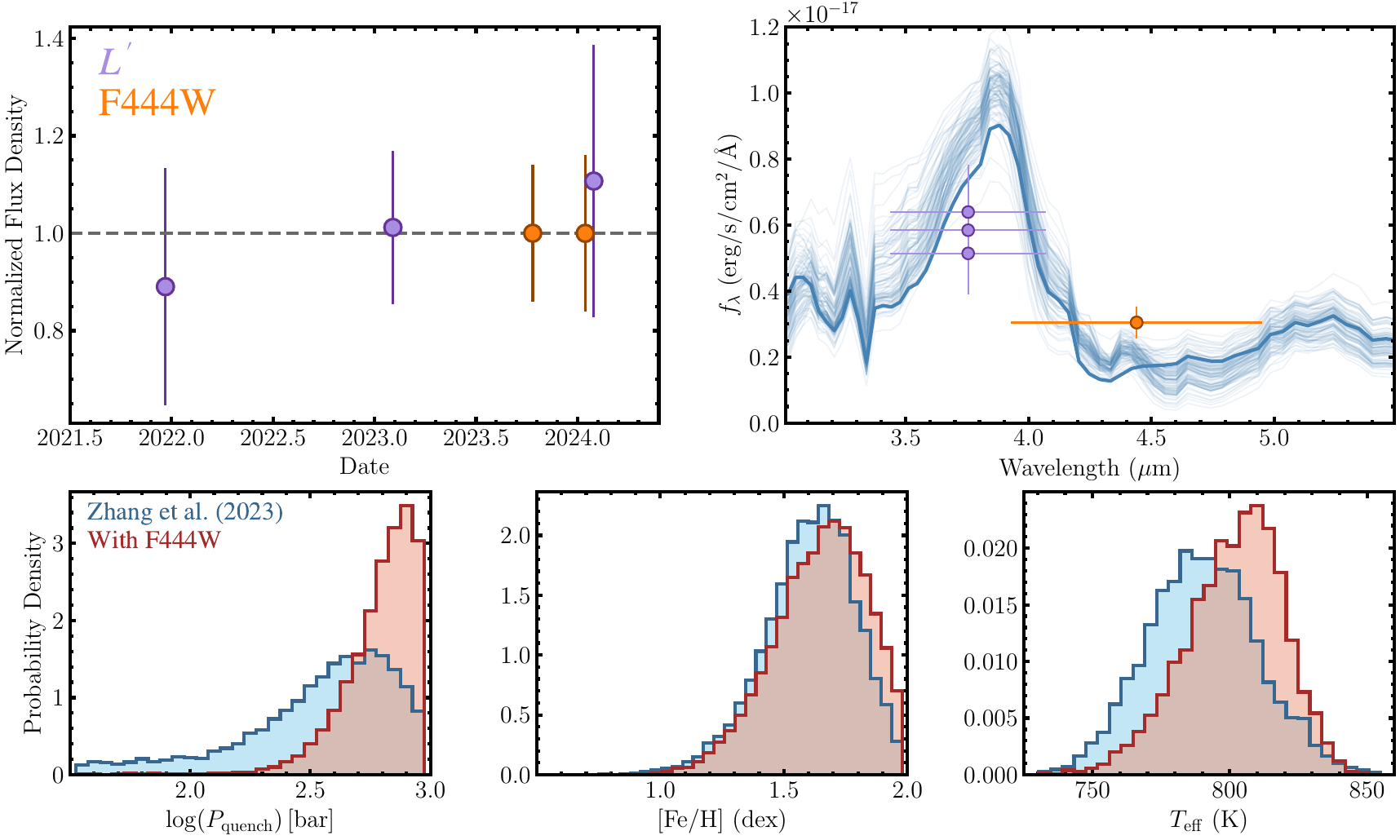}
    \caption{\emph{Top left:} $L'$ (purple) and F444W (orange) flux densities of AF Lep b from December 2021 to January 2024, normalized to the average flux density in each filter. All photometry at these wavelengths is consistent within $1\sigma$ with no evidence for large (${>}20\%$) variability. \emph{Top right:} Retrieved emission spectra of AF Lep b (blue), compared against its $L'$ and F444W photometry. The best-fit model is shown with the thick line, while thin lines denote 100 random emission spectra generated from the retrieved posteriors. \emph{Bottom:} Comparison of retrieved parameters for $\log(P_\mathrm{quench})$ (left), $\mathrm{[Fe/H]}$ (center), and $T_\mathrm{eff}$ (right) between our updated retrieval, which includes F444W and new $L'$ photometry, and the previous equivalent retrieval in \citet{zhangElementalAbundancesPlanets_2023}. $T_\mathrm{eff}$ and $\mathrm{[Fe/H]}$ produce similar posteriors, while the uncertainty on $\log(P_\mathrm{quench})$, which parameterizes disequilibrium chemistry, decreases substantially.
    \label{fig:retrieval_variability}}
\end{figure*}

\subsection{Atmospheric Modeling}
The NIRCam detection of AF Lep b in F444W extends the planet's Spectral Energy Distribution (SED) into the $4{-}5 \, \mathrm{\mu m}$ regime (see Figure \ref{fig:reducedimgs}, bottom row). In this section, we examine the impact of this additional photometry on the inferred atmospheric properties through two approaches. First, we compare the 3--$5 \, \mathrm{\mu m}$ SED against the predictions of grid-based models. We then perform a new atmospheric retrieval within the framework of \citet{zhangElementalAbundancesPlanets_2023}, incorporating all published photometry and spectroscopy of AF Lep b.

\subsubsection{Grid-based Model Comparison \label{sec:forward_models}}
The main spectral features from $3{-}5 \, \mathrm{\mu m}$ for L and T dwarfs are the $\nu_3$ fundamental absorption band of $\mathrm{CH_4}$ from ${\sim}3.0{-}3.8 \, \mathrm{\mu m}$ and the $\nu{=}1{-}0$ fundamental absorption band of CO from ${\sim}4.4{-}5.0\, \mathrm{\mu m}$. The $\mathrm{CO_2}$ $\nu_3$ fundamental band may also contribute to absorption from $4.2{-}4.4 \, \mathrm{\mu m}$. Combining the $L'$ photometry of AF Lep b, which samples $\mathrm{CH_4}$ absorption, with the F444W photometry, which samples CO and $\mathrm{CO_2}$ absorption, thus provides a means to constrain the planet's carbon chemistry and the presence of disequilibrium processes elevating its CO abundance. \citet{derosaDirectImagingDiscovery_2023} performed a similar exercise in the $K$-band, comparing the $K1{-}K2$ color of AF Lep b to other directly imaged planets, finding that AF Lep b has a color intermediate between that of red L-dwarf planets and 51 Eri b. This hints at the beginning of methane absorption in the atmosphere of AF Lep b.

The $L'-\mathrm{F444W}$ color of AF Lep b is $-0.02 \pm 0.16 \, \mathrm{mag}$. The sample of imaged exoplanets with $L'$ and F444W photometry for direct comparison is limited to the mid-to-late L-dwarf planet HIP 65426 b \citep{chauvinDiscoveryWarmDusty_2017}, which has an $L'-\mathrm{F444W}$ color of $-0.23 \pm 0.14 \, \mathrm{mag}$ \citep{cheethamSpectralOrbitalCharacterisation_2019,carterJwstEarlyRelease_2023}, and the early L-dwarf planet $\beta$ Pic b \citep{lagrangeGiantPlanetImaged_2010,bonnefoyPhysicalOrbitalProperties_2014}, which has an $L'-\mathrm{F444W}$ color of $0.11 \pm 0.08 \, \mathrm{mag}$ \citep{currieCombinedVeryLarge_2013,kammererJwsttstHighContrast_2024}. Bluer $L'{-}\mathrm{F444W}$ colors indicate CO absorption, while redder colors point to $\mathrm{CH_4}$ absorption. To further contextualize this spectral slope, we compare the $3{-}5 \, \mathrm{\mu m}$ photometry of AF Lep b against \texttt{Sonora Bobcat} \citep{marleySonoraBrownDwarf_2021} and \texttt{Sonora Elf Owl} \citep{mukherjeeSonoraSubstellarAtmosphere_2024} model atmospheres. The \texttt{Sonora Bobcat} grid is computed assuming chemical equilibrium and cloudless atmospheres, varying metallicity, $\log g$, $T_\mathrm{eff}$, and C/O ratio. \texttt{Sonora Elf Owl} models extend \texttt{Bobcat} models with a self-consistent treatment of disequilibrium chemistry across a wide range of effective temperatures ($275$ to $2400 \, \mathrm{K}$), C/O ratios ($0.22$ to $1.24$), and metallicities ($\mathrm{[M/H]} {=} -1.0$ to $+1.0$). Disequilibrium chemistry from vertical mixing is parametererized by the eddy diffusion coefficient $K_{zz}$ \citep{griffithDisequilibriumChemistryBrown_1999}, where high values of $K_\mathrm{zz}$ indicate strong vertical mixing, while low values of $K_\mathrm{zz}$ imply weaker mixing. 

Figure \ref{fig:combined_analysis} (top) compares the $L{-}\mathrm{F444W}$ color and effective temperature of AF Lep b to \texttt{Bobcat} and \texttt{Elf Owl} models with $\log{g} = 3.5 \, \mathrm{dex}$; $\mathrm{C/O} = 0.458$ (corresponding to solar; \citealt{loddersSolarSystemElemental_2009}); and $\mathrm{[M/H]} = +0.5$, 0, and $-0.5$. The surface gravity is selected to match the inferred $\log{g}$ of AF Lep b of ${\sim}3.7$ dex from \citet{palma-bifaniAtmosphericPropertiesAf_2024} and \citet{zhangElementalAbundancesPlanets_2023}, while the C/O ratio and metallicity enable the inclusion of \texttt{Bobcat} models, which span a smaller range of metallicities and have limited coverage for non-solar C/O ratios. Within the \texttt{Elf Owl} models, $K_{zz}$ varies from $10^2$ to $10^9 \, \mathrm{cm^2 \, s^{-1}}$. For the effective temperature of AF Lep b, we adopt a nominal value of $800 \pm 100 \, \mathrm{K}$ based on effective temperatures from forward modeling (${\sim}750 \, \mathrm{K}$; \citealt{palma-bifaniAtmosphericPropertiesAf_2024}) and retrieval (${\approx} 800 \, \mathrm{K}$; \citealt{zhangElementalAbundancesPlanets_2023} and this work) analyses. As expected, models with lower values of $\log{K_{zz}}$ corresponding to less vigorous vertical mixing have bluer $L'{-}\mathrm{F444W}$ colors, with equilibrium models at the far blue end. This reflects the increase in CO abundance with more vigorous mixing at these effective temperatures, where faster mixing inhibits the conversion of CO to $\mathrm{CH_4}$. However, as discussed by \citet{mukherjeeSonoraSubstellarAtmosphere_2024}, increased $K_\mathrm{zz}$ is degenerate with increased atmospheric metallicity in the 3--$5 \, \mathrm{\mu m}$ regime, which also increases CO abundance relative to $\mathrm{CH_4}$. This is seen in Figure 2 (top), where higher-metallicity models generally produce lower $L'{-}\mathrm{F444W}$ colors at a given $T_\mathrm{eff}$ and $\log{K_{zz}}$.

The blue $L'{-}\mathrm{F444W}$ color of AF Lep b is inconsistent with sub-solar metallicity models, consistent within $1\sigma$ of solar-metallicity \texttt{Elf Owl} models with the most vigorous mixing $\log{K_{zz}} = 9$, and consistent with $\mathrm{[M/H]} = +0.5$ models with relatively high mixing levels ($\log{K_{zz}} = 7$ or $\log{K_{zz}} = 8$). Figure \ref{fig:combined_analysis} (bottom) compares model spectra against the $L'$ and F444W photometry of AF Lep b. Each model spectrum is anchored to the weighted mean of the $L'$ photometry, with synthesized F444W photometry shown in open circles. The left panel shows models varying $\log{K_\mathrm{zz}}$ and metallicity, while the right panel shows models varying metallicity with $\log{K_\mathrm{zz}} = 2$. In general, the F444W photometry is best reproduced by models with high values of $\log K_{zz}$ and slightly enhanced or solar metallicites ($\mathrm{[M/H]} = 0$ or $\mathrm{[M/H]} = +0.5$) or lower values of $\log K_{zz}$ with super-solar metallicities ($\mathrm{[M/H]} = +1$). Note that \texttt{Bobcat} models do not extend to metallicities of $+1.0 \, \mathrm{dex}$. However, even with $\log{K_\mathrm{zz}} = 2$ \texttt{Elf Owl} models, the synthesized F444W photometry remains above the observed values. This is clear evidence for the presence of disequilibrium carbon chemistry in the atmosphere of AF Lep b. 

However, degeneracies remain between the metallicity enhancement and the rate of vertical mixing driving disequilibrium chemistry, both of which our observations support. The $4.3 \, \mathrm{\mu m}$ $\mathrm{CO_2}$ feature can help to break this degeneracy, as $\mathrm{CO_2}$ abundance is enhanced by increased metallicity, but is only weakly enhanced by disequilibrium processes \citep{mukherjeeSonoraSubstellarAtmosphere_2024}. Follow-up medium-band JWST photometry or $M$-band photometry that only samples $\mathrm{CO_2}$ or CO absorption may enable more robust constraints to be placed on atmospheric metallicity and $\log{K_{zz}}$ within this spectral region. Another limitation of this analysis is that \texttt{Bobcat} and \texttt{Elf Owl} models are cloudless. Clouds likely affect the spectrum of AF Lep b, although the impact at these wavelengths is significantly less than at shorter wavelengths and smaller in this regime than the presence of disequilibrium chemistry (see Figure 6 of \citealt{zhangElementalAbundancesPlanets_2023}).

\subsubsection{Updated Retrieval}
To further assess the impact of the new $4.4\, \mathrm{\mu m}$ photometry, we perform an updated retrieval following \citet{zhangElementalAbundancesPlanets_2023}. Within this framework, \texttt{petitRADTRANS} \citep{mollierePetitradtransPythonRadiative_2019} is used to perform chemically-consistent retrievals. A novel temperature-pressure (T-P) profile parameterization is adopted in which the T-P profile is anchored to six temperature gradients evenly spaced across the logarithmic pressure scale. Priors on the temperature gradients informed by forward model T-P profiles are adopted to enforce radiative convective equilibrium and avoid retrievals preferring more isothermal, less cloudy T-P profiles than expected \citep[e.g.,][]{molliereRetrievingScatteringClouds_2020,whitefordRetrievalStudyCool_2023}. Disequilibrium chemistry follows a different parameterization than the forward models in Section \ref{sec:forward_models}: the logarithmic quench pressure $\log(P_\mathrm{quench})$ is included as a free parameter. For pressures below $P_{\mathrm{quench}}$, the abundances of $\mathrm{H_2 O}$, CO, and $\mathrm{CH_4}$ are set to their abundances at $P_\mathrm{quench}$ \citep[e.g.,][]{zahnleMethaneCarbonMonoxide_2014}. In this way, $P_\mathrm{quench}$ represents the amount of the atmosphere affected by disequilibrium chemistry, with higher values of $P_\mathrm{quench}$ indicating a larger extent of the atmospheric profile with abundances impacted by vertical mixing. See \citet{zhangElementalAbundancesPlanets_2023} and \citet{mollierePetitradtransPythonRadiative_2019} for additional details on the retrieval framework.

We perform a retrieval incorporating all photometry and spectroscopy of AF Lep b from \citet{fransonAstrometricAccelerationsDynamical_2023a}, \citet{derosaDirectImagingDiscovery_2023}, \citet{mesaAfLepLowest_2023}, and this work. Free parameters include the temperature at the bottom layer of the pressure profile $T_\mathrm{bottom}$; T-P gradients at six pressure levels $(d \ln T / d \ln P)_{1 . . . 6}$; metallicity (defined as $\mathrm{[Fe/H]}$); C/O ratio; $\log(P_\mathrm{quench}) \, \mathrm{[bar]}$; mass fractions at the base pressure ($\log(X_0)$) and sedimentation efficiencies ($f_\mathrm{sed}$) for $\mathrm{MgSiO_3}$, Fe, and KCl clouds; $\log{K_\mathrm{zz}}$; the width of the log-normal cloud particle size distribution $\sigma_g$; $\log(g)$; radius $R$; and flux offsets for the SPHERE/IFS spectra from \citet{derosaDirectImagingDiscovery_2023} and \citet{mesaAfLepLowest_2023} ($\Delta f_\mathrm{D23}$ and $\Delta f_\mathrm{M23}$, respectively). Note that here, disequilibrium chemistry is not coupled to $\log{K_\mathrm{zz}}$: the eddy diffusion coefficient is solely used to set the cloud particle size distribution, while $\log(P_\mathrm{quench})$ determines non-equilibrium abundances \citep{molliereRetrievingScatteringClouds_2020}. We follow \citet{zhangElementalAbundancesPlanets_2023} in setting priors for these free parameters. The priors for all parameters other than $\log(g)$ and $R$ can be found in Table 5 of \citet{zhangElementalAbundancesPlanets_2023}. For $R$, a uniform prior from $1.20 \, R_\mathrm{Jup}$ to $1.55 \, R_\mathrm{Jup}$ is applied, informed by evolutionary models \citep{zhangElementalAbundancesPlanets_2023}. This avoids the small radius problem that has been a recurrent issue with retrievals \citep[e.g.,][]{kitzmannHeliosr2NewBayesian_2020,lueberRetrievalStudyBrown_2022}. The $\log(g)$ prior is inferred through the radius prior and dynamical mass from \citet{zhangElementalAbundancesPlanets_2023} of $2.8 \pm 0.6 \, M_\mathrm{Jup}$. These constrained priors were also applied for some of the retrievals presented in \citet{zhangElementalAbundancesPlanets_2023}.

The updated retrieval produces the following posteriors for key parameters: 
$\mathrm{[Fe/H]} {=} {1.67}_{-0.21}^{+0.17} $, $\mathrm{C/O} {=} {0.65}_{-0.09}^{+0.07} $, $\log(L_\mathrm{bol}/L_\odot) {=} {-}5.191 {\pm} 0.023 $, $\log(g) {=} {3.65}_{-0.05}^{+0.04} $, $R {=} {1.27}_{-0.05}^{+0.07} \, R_\mathrm{Jup}$, $T_\mathrm{eff} {=} {803}_{-19}^{+15} \, \mathrm{K}$, and $\log(P_\mathrm{quench}) {=} {2.81}_{-0.26}^{+0.12} $. These results can be directly compared against the rightmost column of Table 6 in \citet{zhangElementalAbundancesPlanets_2023}. All parameters are consistent within $1\sigma$ with the prior retrieval. Figure \ref{fig:retrieval_variability} (bottom) compares the distributions of $\log(P_\mathrm{quench})$, $\mathrm{[Fe/H]}$, and $T_\mathrm{eff}$ between the two retrievals. The $T_\mathrm{eff}$ and $\mathrm{[Fe/H]}$ posteriors are similar, with our updated retrieval still finding an enhanced atmospheric metallicity. The $\mathrm{[Fe/H]}$ posterior corresponds to a metallicity enrichment of $Z_\mathrm{pl}/Z_* = 50^{+60}_{-30}$. Here, we compute $Z_\mathrm{pl}/Z_*$ from Equations 19 and 20 of \citet{nasedkinFourofakindComprehensiveAtmospheric_2024} following the same assumptions as \citet{thorngrenConnectingGiantPlanet_2019} for a host-star metallicity of $-0.27 \pm 0.32$ \citep{zhangElementalAbundancesPlanets_2023}. This value is higher than the predicted metallicity enrichment for planets with the mass of AF Lep b; the empirical mass-metallicity relation from \citet{thorngrenMassMetallicityRelation_2016} predicts $Z_\mathrm{pl}/Z_* = 6.1^{+1.2}_{-1.1}$ for a $2.8^{+0.6}_{-0.5} \, M_\mathrm{Jup}$ planet. Note, though, that forward modeling analysis of AF Lep b has produced more modest metallicity enhancements ($\mathrm{[Fe/H]} {\approx} +0.6$; \citealt{palma-bifaniAtmosphericPropertiesAf_2024}). The main difference between the retrievals is that the $\log(P_\mathrm{quench})$ posterior is better constrained within the updated retrieval; the previous retrieval found $\log{(P_\mathrm{quench})} = 2.5^{+0.3}_{-1.1}$, while the updated retrieval produces a posterior of ${2.81}_{-0.26}^{+0.12}$. Overall, extending the SED of AF Lep b to $4.4\, \mathrm{\mu m}$ enables more precise constraints on the interior dynamics of the atmosphere of AF Lep b through sampling the carbon disequilibrium chemistry. The new photometry also affirms previous evidence for enhanced atmospheric metallicity.

\subsection{Variability}
Rotationally-modulated near-infrared variability is ubiquitous among isolated brown dwarf- and planetary-mass objects. L and T dwarfs exhibit variability with typical amplitudes of 0.2--5\% from $1{-}5\, \mathrm{\mu m}$ \citep[e.g.,][]{artigauPhotometricVariabilityT2_2009,metchevWeatherOtherWorlds_2015,billerSimultaneousMultiwavelengthVariability_2018,zhouSpectralVariabilityVhs_2020}, although amplitudes up to 38\% in $J$ band have been observed \citep{bowlerStrongNearinfraredSpectral_2020,zhouRoaringStormsPlanetarymass_2022}. The primary source of this variability, at least for late-L- and T-type objects, is likely heterogeneous clouds \citep[e.g.,][]{apaiHstSpectralMapping_2013,morleySpectralVariabilityPatchy_2014,vosPatchyForsteriteClouds_2023,mccarthyMultiplePatchyCloud_2024}. In line with this interpretation, variability amplitudes and occurrence rates are higher at young ages \citep{vosLetGreatWorld_2022}, the L/T transition \citep{radiganStrongBrightnessVariations_2014}, and equator-on viewing angles \citep{vosViewingGeometryBrown_2017}. These trends bode well for the potential to use photometric variability to characterize directly imaged planets. Moreover, Jupiter exhibits disk-averaged variability of ${\sim}4\%$ in the optical and ${\sim}20\%$ at $4.7 \, \mathrm{\mu m}$ \citep{gelinoVariabilityUnresolvedJupiter_2000,geRotationalLightCurves_2019}. However, despite several attempts \citep{apaiHighcadenceHighcontrastImaging_2016,billerHighcontrastSearchVariability_2021,wangAtmosphericMonitoringPrecise_2022}, variability has yet to be detected for resolved, imaged exoplanets on closer-in (${<}100\, \mathrm{au}$) orbits, where photometric precision is limited by PSF subtraction.

The two epochs of F444W photometry and three epochs of $L'$ photometry of AF Lep b do not show evidence of variability at the 10\% and 20\% levels, respectively. Figure \ref{fig:retrieval_variability} (top left) displays the flux densities of the $L'$ and F444W photometry of AF Lep b, normalized to the average flux density for each filter. All photometry is consistent within $1 \sigma$. The F444W photometry has an average flux density of $(3.05 \pm 0.32) \times 10^{-18} \, \mathrm{erg/s/cm^2/\text{\AA}}$. The two epochs separated by three months are identical within their precisions, differing by $0.0 \pm 0.7 \, \mathrm{erg/s/cm^2/\text{\AA}}$ or $0.00 \pm 0.23 \, \mathrm{mag}$. Here, the 1.5\% absolute flux calibration uncertainty is not included, as a NIRCam flux offset would have a uniform effect on both photometric points. The average $L'$ flux density is $(5.8 \pm 0.6) \times 10^{-18} \, \mathrm{erg/s/cm^2/\text{\AA}}$. Each $L'$ point, spanning a baseline of two years, is consistent within $1 \sigma$ of the average. The ratio of the standard deviation of all photometric points to the mean photometric uncertainty, $\sigma /\overline{\sigma}_\mathrm{meas}$, offers a way to compare the significance level of potential variability signals.  Here, $\sigma/\overline{\sigma}_\mathrm{meas} \gg 1$ indicates variability, while $\sigma/\overline{\sigma}_\mathrm{meas} \lesssim 1$ indicates no evidence for variability. This ratio is 0 for F444W and 0.54 for $L'$.

\section{Summary}
In this work, we presented JWST/NIRCam imaging of the $3 \, M_\mathrm{Jup}$ planet AF Lep b. AF Lep b is recovered at a significance of $9.6\sigma$ and $8.7\sigma$ in two epochs of F444W NIRcam imaging. This extends the SED of this recently discovered planet into the 4--$5\, \mathrm{\mu m}$ regime. We also presented new Keck/NIRC2 $L'$ imaging of AF Lep b. Our main conclusions are summarized below:
\begin{itemize}
    \item AF Lep b is relatively faint in F444W, with an average flux density 53\% of its average flux density in $L'$, indicating significant CO absorption. At the effective temperature of AF Lep b, this points to the presence of disequilibrium carbon chemistry driven by vertical mixing. Enhanced atmospheric metallicity also increases CO abundance, although at $\mathrm{[M/H]} {=} +0.5$, only forward models with vigorous vertical mixing ($\log{K_\mathrm{zz}} > 4$) are consistent with the observed 3--$5\, \mathrm{\mu m}$ photometry.
    \item An updated retrieval within the \citet{zhangElementalAbundancesPlanets_2023} framework produces consistent posteriors with the previous retrieval for most parameters. The uncertainty on the quench pressure ($\log(P_\mathrm{quench})$), which is used to parameterize disequilibrium chemistry, decreases significantly. An enhancement in the atmospheric metallicity is still found. Overall, the F444W photometry provides clear evidence for disequilibrium chemistry within the atmosphere of AF Lep b while affirming previous evidence for enhanced atmospheric metallicity.
    \item No additional planets on wider orbits are detected around AF Lep. Our F444W contrast curve places deep upper limits on the mass of additional companions in the system, ruling out $1.9\, M_\mathrm{Jup}$ planets beyond 0\farcs5, $1.1 \, M_\mathrm{Sat}$ planets beyond $1^{\prime\prime}$, and $2.8 \, M_\mathrm{Nep}$ planets beyond $2\farcs5$. This largely rules out the scenario of an additional giant planet sculpting the debris disk around AF Lep.
\end{itemize}

The successful imaging of AF Lep b with JWST at its current separation of ${\approx}320\, \mathrm{mas}$ and F444W contrast of ${\approx} 10 \, \mathrm{mag}$ is an exciting technical achievment. Alongside the recent MIRI \citep{boccalettiImagingDetectionInner_2023} detection of HR 8799 e at a separation of $350\, \mathrm{mas}$, our recovery of AF Lep b demonstrates that JWST coronagraphy is outperforming pre-launch expectations at close-in separations.

\section{Acknowledgements}
We greatly appreciate the efforts of the JWST mission operations team for rapidly scheduling this DD program. We thank Jarron Leisenring and Julien Girard for helpful conversations on the throughput of the occulting mask. K.F. acknowledges support from the National Science Foundation Graduate Research Fellowship Program under Grant No. DGE 2137420. B.P.B. acknowledges support from the National Science Foundation grant AST-1909209, NASA Exoplanet Research Program grant 20-XRP20$\_$2-0119, and the Alfred P. Sloan Foundation. T.D.P is supported by a UKRI/EPSRC Stephen Hawking Fellowship. This work is based on observations made with the NASA/ESA/CSA James Webb Space Telescope. The data were obtained from the Mikulski Archive for Space Telescopes at the Space Telescope Science Institute, which is operated by the Association of Universities for Research in Astronomy, Inc., under NASA contract NAS 5-03127 for JWST. These observations are associated with program \#4558. Support for program \#4558 was provided by NASA through a grant from the Space Telescope Science Institute, which is operated by the Association of Universities for Research in Astronomy, Inc., under NASA contract NAS 5-03127. The JWST data presented in this article were obtained from the Mikulski Archive for Space Telescopes (MAST) at the Space Telescope Science Institute. The specific observations analyzed can be accessed via \dataset[doi: 10.17909/eag5-4157]{https://doi.org/10.17909/eag5-4157}.
This research has made use of the VizieR catalogue access tool, CDS, Strasbourg, France (DOI: 10.26093/cds/vizier). The original description of the VizieR service was published in 2000, A\&AS 143, 23. This publication makes use of data products from the Wide-field Infrared Survey Explorer, which is a joint project of the University of California, Los Angeles, and the Jet Propulsion Laboratory/California Institute of Technology, funded by the National Aeronautics and Space Administration. Based on observations made with ESO Telescopes at the La Silla Paranal Observatory under programme ID 2111.C-5021(B). This work was supported by a NASA Keck PI Data Award, administered by the NASA Exoplanet Science Institute. Data presented herein were obtained at the W. M. Keck Observatory from telescope time allocated to the National Aeronautics and Space Administration through the agency's scientific partnership with the California Institute of Technology and the University of California. The Observatory was made possible by the generous financial support of the W. M. Keck Foundation.

The authors wish to recognize and acknowledge the very significant cultural role and reverence that the summit of Maunakea has always had within the indigenous Hawaiian community. We are most fortunate to have the opportunity to conduct observations from this mountain.

\software{\texttt{pyKLIP} \citep{wangPyklipPsfSubtraction_2015}, \texttt{spaceKLIP} \citep{kammererPerformanceNearinfraredHighcontrast_2022}, \texttt{pysynphot} \citep{stscidevelopmentteamPysynphotSyntheticPhotometry_2013}, \texttt{webbpsf} \citep{perrinSimulatingPointSpread_2012,perrinUpdatedPointSpread_2014}, \texttt{ccdproc} \citep{craigAstropyCcdprocV1_2017}, \texttt{photutils} \citep{bradleyAstropyPhotutilsV0_2019}, \texttt{astropy} \citep{astropycollaborationAstropyCommunityPython_2013,astropycollaborationAstropyProjectBuilding_2018,astropycollaborationAstropyProjectSustaining_2022}, \texttt{pandas} \citep{mckinneyDataStructuresStatistical_2010}, \texttt{matplotlib} \citep{hunterMatplotlib2dGraphics_2007}, \texttt{numpy} \citep{harrisArrayProgrammingNumpy_2020}, \texttt{scipy} \citep{virtanenScipyFundamentalAlgorithms_2020}, \texttt{emcee} \citep{foreman-mackeyEmceeMcmcHammer_2013}, \texttt{corner} \citep{foreman-mackeyCornerPyScatterplot_2016}}

\appendix

\begin{figure*}
    \centering
    \includegraphics[width=1\textwidth]{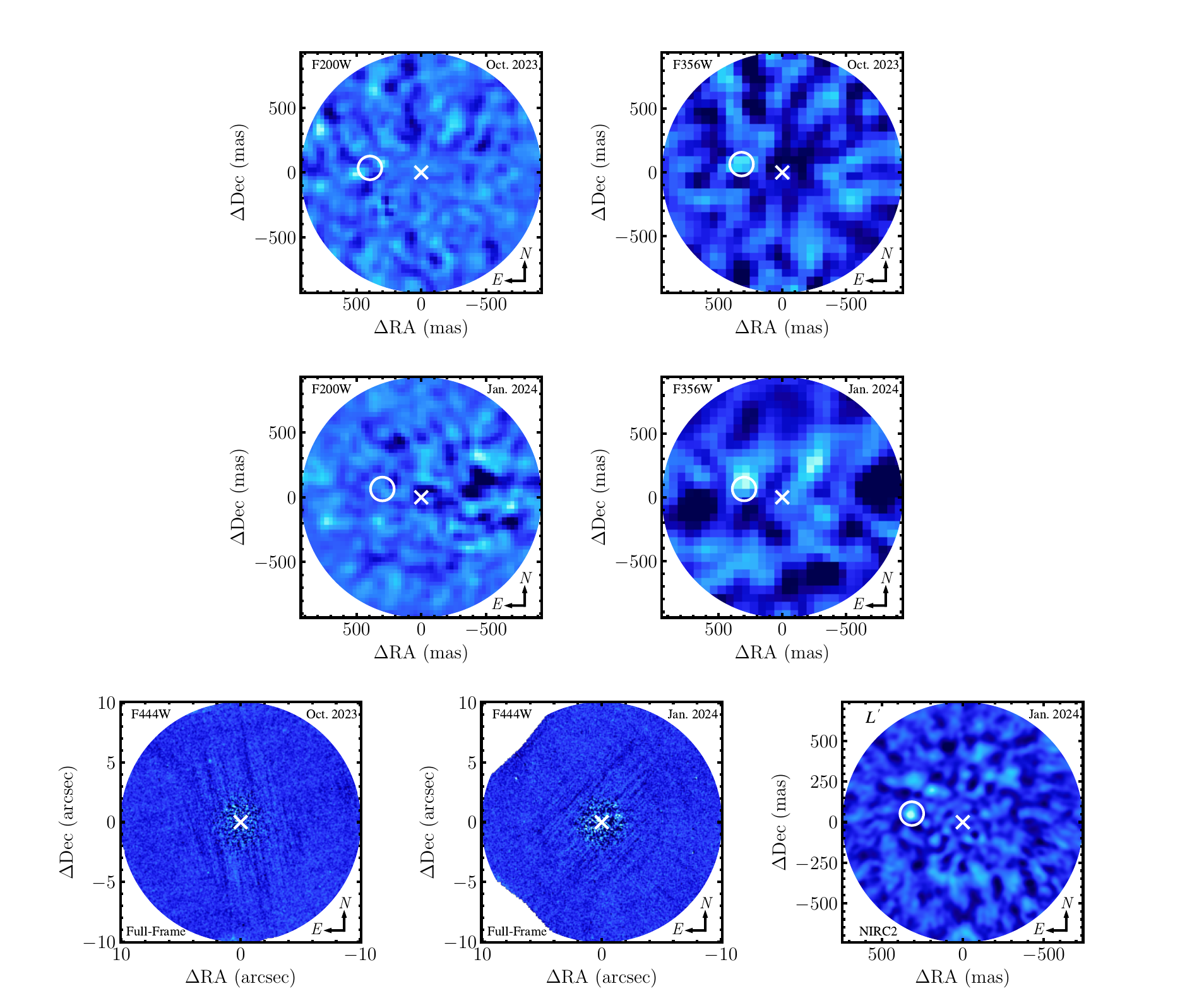}
    \caption{\emph{Top and Middle:} F200W (left) and F356W (right) imaging of AF Lep. AF Lep b is not robustly recovered (${>}5\sigma$) in either filter or epoch. Like the F444W imaging in Figure \ref{fig:reducedimgs}, each image is convolved with a Gaussian filter with a standard deviation of 1 pixel. The circles denote the expected position of AF Lep b at each epoch. The misalignment between the coronagraphic mask and the host star for each dataset is incorporated into this prediction. \emph{Bottom:} The left and center panels show full-frame F444W imaging of AF Lep for each sequence. No significant point sources are seen at wider separations in this imaging. The artifact to the NE of the star in the October 2023 imaging is a cosmic ray artifact that does not appear in the January 2024 imaging. The right panel shows our new Keck/NIRC2 $L'$ imaging of AF Lep b. To smooth over pixel-to-pixel noise, the image is convolved with a Gaussian filter with a standard deviation of 1.5 pixels. The circle highlights the expected position of the planet. AF Lep b is recovered at a S/N of $6.6\sigma$. 
    \label{fig:remaining_imaging}}
\end{figure*}

\section{Additional JWST and Keck/NIRC2 Imaging\label{sec:all_imgs}} In this section, we present additional reduced frames from the JWST and Keck/NIRC2 datasets not shown in Figure \ref{fig:combined_analysis}. Figure \ref{fig:remaining_imaging} displays the F200W and F356W imaging in which AF Lep b is not detected at a significant level (top and middle rows), full-frame F444W imaging (bottom row, center and left), and the January 2024 Keck/NIRC2 $L'$ imaging (bottom row, right). A speckle appears in both F356W datasets at the expected position of AF Lep b. However, the S/N of these speckles ($3.7 \sigma$ and $3.2\sigma$ in October 2023 and January 2024, respectively) falls below our nominal cutoff ($5\sigma$) for claiming a detection. No significant sources are seen in the F200W imaging. The effect of the pointing error in the January 2024 data, which is in the SE direction from the mask center, can be seen in the January 2024 F200W imaging. No significant point sources are seen at wider separations in the full-frame F444W imaging. The source at $(\Delta \mathrm{RA}, \ \Delta \mathrm{Dec}) \approx (1\farcs2, \ 5^{\prime\prime})$ in the October imaging is a cosmic ray ``snowball'' artifact \citep[see e.g.,][]{bagleyCeersEpochNircam_2023} that does not appear in the January observations.

\bibliography{references}{}
\bibliographystyle{aasjournal}
\newpage

\end{document}